\documentclass[]{iopart}

\usepackage{graphicx}
\usepackage{dcolumn}
\usepackage{bm}
\usepackage{amssymb}
\usepackage{nicefrac}
\usepackage{color}
\usepackage{citesort}

\makeatletter
\g@addto@macro\@floatboxreset\centering
\makeatother

\begin{document}

\author{Wolf B Dapp$^{1}$, Nikolay Prodanov$^{1,2}$ and Martin H M\"user$^{1,2}$}
\eads{\mailto{w.dapp@fz-juelich.de},\mailto{martin.mueser@mx.uni-saarland.de}}
\address{$^1$ J\"ulich Supercomputing Centre, Institute for Advanced Simulation, FZ J\"ulich, 52425 J\"ulich, Germany}
\address{$^2$ Dept. of Materials Science and Engineering, Universit\"at des Saarlandes, 66123 Saarbr\"ucken, Germany}

\title{
Systematic analysis of Persson's contact mechanics theory of randomly 
rough elastic surfaces
}


\begin{abstract} 
We systematically check explicit and implicit assumptions of Persson's 
contact mechanics theory. 
It casts the evolution of the pressure distribution ${\rm Pr}(p)$ with increasing
resolution of surface roughness as a diffusive process, in which resolution 
plays the role of time. 
The tested key assumptions of the theory are:
(a) the diffusion coefficient is independent of pressure $p$,
(b) the diffusion process is drift-free at any value of $p$,
(c) the point $p=0$ acts as an absorbing barrier, i.e., once a point falls out
of contact, it never reenters again,
(d) the Fourier component of the elastic energy is only populated if the appropriate
wave vector is resolved, and
(e) it no longer changes when even smaller wavelengths are resolved.
Using high-resolution numerical simulations, we quantify deviations from these 
approximations and find quite significant discrepancies in some cases.
For example, the drift becomes substantial for small values of
$p$, which typically represent points in real space close to a contact line.
On the other hand, there is a significant flux of points reentering contact. 
These and other identified deviations cancel each other to a large degree, 
resulting in an overall excellent description for contact area, contact geometry, 
and gap distribution functions.
Similar fortuitous error cancellations cannot be guaranteed under different 
circumstances, for instance when investigating rubber friction. 
The results of the simulations may provide guidelines for a systematic 
improvement of the theory.
\end{abstract}

\maketitle

\section{Introduction}

Most natural and industrial solids have rough surfaces, with roughness that is 
close to self-affine and fractal, and spanning several orders of magnitude in 
spatial scale. 
Persson theory~\cite{Persson01,Persson06,Almqvist11} has been shown to describe 
the contact mechanics of such surfaces quite well.
It has been extended to also describe various other properties and phenomena, 
such as adhesion~\cite{Kendall01,Persson01adhes,Lorenz13jpcm},
plasticity~\cite{Persson06,Aifantis87},
contact stiffness~\cite{Campana11,Pastewka13},
leakage~\cite{Persson08JPCM,Lorenz10EPJE,Persson12EPJE,DappEtAl2012PRL},
squeeze-out~\cite{Lorenz10EPJE2}, and mixed lubrication~\cite{Persson09JPCM,PerssonScaraggi2011,ScaraggiCarbonePersson2011}.

Persson theory builds on quite simple and elegant statistical premises about 
how the pressure distribution ${\rm Pr}(p,\zeta)$ changes as the ``magnification'' $\zeta$ is increased, 
i.e., sinusoidal roughness is gradually and systematically added to an initially 
flat, semi-infinite surface in contact with a counter body. 
In its original formulation~\cite{Persson01}, the only input into the theory are experimentally
measurable quantities, in particular the power spectrum of the surface 
roughness, the effective modulus, and the external load.
In recent works~\cite{Yang08JPCM}, a ``fudge parameter'' of order unity is introduced.
Its purpose is to make the theory reflect more accurately the relation between
displacement and elastic energy when contact is partial. 
While Persson theory has been tested numerous times and shown to describe many 
interfacial properties quite accurately, only the final results of the theory
have been under scrutiny.
So far, no quantitative analyses have been reported in the literature  
to what extent the assumptions entering the theory hold and to what degree 
(fortuitous) cancellation of errors may be responsible for its success.
Manners and Greenwood~\cite{MannersGreenwood2006} raise some concerns, mainly
with regards to the boundary conditions employed, but fail to investigate to what 
measure the assumptions influence the results.

This paper quantitatively investigates the main underlying assumptions of 
Persson theory, and quantifies the error each assumption introduces. 
We do this with high-resolution numerical simulations using the GFMD 
method~\cite{Campana06,Kong09}. 
There are no adjustable parameters besides those characterizing the rough surface 
and the ratio of pressure and elastic modulus.
The assumptions of an ideally-elastic, semi-infinite half-plane 
are shared by Persson theory. 
We can test each assumption entering the theory individually and assess its effect, 
which may provide guidelines as to how to correct the theory in the future. 

The approach pursued in this work is to solve numerically the contact mechanics
problem by sequentially increasing the magnification.
During each step of including more small-scale details into the simulation, 
we measure the detailed evolution of the system, e.g., we compute the transition 
probability ${\rm Pr}(p, \zeta+\Delta \zeta \vert p',\zeta)$.
It states the likelihood that the pressure at a given interface point in real space
changes from $p'$ to $p$ as the magnification is increased from 
$\zeta$ to $\zeta + \Delta \zeta$.
Another central observable is how the elastic energy is distributed
among different modes (in Fourier space) as $\zeta$ changes. 
These results are then compared to pertinent expressions in Persson theory. 
Analysis of individual modes provides additional information beyond previous tests 
of the theory that only analyzed integrated properties, for example relative contact 
area~\cite{Hyun04,Campana07,Putignano12,ProdanovDappMueser2014a,YastrebovEtAl2014}, the mean 
gap~\cite{Akarapu11,Almqvist11,Putignano12,ProdanovDappMueser2014a}, the contact stiffness as a 
function of the applied pressure~\cite{Akarapu11,Campana11},
adhesion~\cite{PastewkaRobbins2014}, or the correlation 
functions of contact and pressure~\cite{Persson08JPCMb,Campana08}.

This work is structured as follows: 
in Section~\ref{sec:methods} we summarize key ingredients of Persson theory,
including its main assumptions we set out to test as well as the numerical 
methods we use.
Section~\ref{sec:results} presents the results of our tests. 
We discuss our findings in Section~\ref{sec:conclusions}.

\section{Methods}\label{sec:methods}

In this section, we summarize the  key ingredients of Persson theory 
and the numerical methods we use.
Many controlled approximations enter both Persson theory and the numerical solutions
in a similar fashion.
Thus, our analysis only pertains to the accuracy of the
solution of the idealized contact mechanics model and not to the accuracy of the
idealizations themselves.

The idealizations used in this work are: 
The small-slope approximation, neglect of lateral displacements, 
linear elasticity, semi-infinite bodies, hard-wall interactions, and absence
of adhesion, although the latter can be added to both theory and simulations~\cite{Mueser2014a}.
In addition, we assume self-affine surface spectra, whose height profiles can be 
characterized as colored noise.
The statistical properties of our idealized surfaces are defined by their Hurst 
exponent $H$, cutoff (or roll-off) wave numbers limiting the power law behavior 
at large and small wavelengths, respectively, and a prefactor~\cite{Persson06}.

As pointed out recently in a dimensional analysis~\cite{ProdanovDappMueser2014a},
systems with a cut-off at large and small wavelengths ---
which we consider here --- are fully defined by a small set
of dimensionless numbers: 
(i) a dimensionless pressure $\tilde{p}_0 = p_0/E^*\bar{g}$, where $p_0$ is the 
dimensional pressure, $E^*$ the effective modulus, and $\bar{g}$ the root-mean-square
gradient of the surface,
(ii) the Hurst exponent $H$, and 
(iii) the ratio of the two cut-offs at short and long wavelengths, 
i.e., $\epsilon_{\rm f} = \lambda_{\rm s}/\lambda_{\rm l}$. 
In addition, one may consider
(iv) the ratio of system size ${\cal L}$ and $\lambda_{\rm l}$, which, however, is 
only relevant at very small loads~\cite{Pastewka13}, and

(v), in numerical simulations, the ratio of lattice discretization $a$ and $\lambda_{\rm s}$, which we try
to keep small enough to approach the continuum limit sufficiently well.

\subsection{Persson theory}

The contact mechanics theory by Persson has been summarized several 
times~\cite{Persson01,Persson05JPCM,Persson06}, also in 
a previous work~\cite{ProdanovDappMueser2014a}.
Here we focus on its details related to the assumptions we are going to verify.

Assume we know the pressure distribution in a contact, whose spatial features
are  resolved up to a magnification of $\zeta$, i.e., the spectrum of the
surfaces is limited to wavevectors magnitudes $q_{\rm l} \le q \le \zeta q_{\rm l}$,
where $q_{\rm l} = 2\pi/\lambda_{\rm l}$. 
We could predict how the distribution changes with increasing $\zeta$ 
if we knew the transition probability 
${\rm Pr}(p, \zeta+\Delta \zeta \vert p',\zeta)$, which, as stated in the 
introduction, specifies the likelihood that the pressure at a given point in real 
space changes from $p'$ to $p$ as the magnification is increased from
$\zeta$ to $\zeta + \Delta \zeta$.
By definition of the transition probability, one would obtain
\begin{equation}
{\rm Pr}(p, \zeta + \Delta \zeta) = 
\int {\rm d}p'\,
{\rm Pr}(p, \zeta+\Delta \zeta \vert p',\zeta)
{\rm Pr}(p', \zeta).
\end{equation}
The starting point of Persson theory is an approximation
to this transition probability according to:
\begin{equation}
{\rm Pr}(p, \zeta+\Delta \zeta \vert p',\zeta) \approx
\frac{1}{\sqrt{2\pi\Delta p^2}}\exp\left\{- \frac{(p-p')^2}{2\Delta p^2}\right\}
\label{eq:1Gaussian}
\end{equation}
with
\begin{equation}
\Delta p^2 = \sum_{\zeta q_{\rm l} \le \vert{\bf q}\vert < (\zeta+\Delta \zeta) q_{\rm l}} 
\left(\frac{qE^*}{2}\right)^2 \vert \tilde{h}({\bf q}) \vert^2,
\label{eq:broadening}
\end{equation}
where the $\tilde{h}({\bf q})$ denote the Fourier coefficients of the 
(combined) surface height. 
The broadening of pressure $\Delta p$ is motivated from the exact expression
for the broadening that would hold if an infinitesimally small, single-wave
height fluctuation were added to an otherwise perfectly smooth  interface
under a finite normal pressure. 
It appears that it suffices to use a Gaussian in place of the (unknown) true broadening 
function, because --- owing to the central limit theorem --- the detailed shape of 
transition probabilities should become irrelevant 
when they are applied repeatedly a large number of times.

One problem of the Gaussian transition probability is that it also allows negative 
pressures.
This  undesired property can be avoided by interpreting the broadening of 
the pressure distribution function in terms of a diffusive process:
each point in the interface represents a walker, the pressure could be
seen as its ``random location'', and magnification plays the role of time.
For example, as magnification goes on, the local pressure would be expected
to mount when the local height increases relative to
some neighborhood with greater $\zeta$, while it would diminish in the opposite case. 

The boundary condition of non-negative pressures not being allowed in the
given context can be implemented within this interpretation
by assuming that each walker hitting  the $p=0$ boundary  
gets absorbed into  it, that is, the walker gets lost to noncontact. 
The idea can be realized formally by subtracting a mirror Gaussian from the original 
Gaussian in \eref{eq:1Gaussian} so that
\begin{eqnarray}
\fl{\rm Pr}(p>0, \zeta+\Delta \zeta \vert p',\zeta)  = &
\frac{1}{\sqrt{2\pi\Delta p^2}} \times
\nonumber\\
& \left[ \exp\left\{- \frac{(p-p')^2}{2\Delta p^2}\right\} 
-  \exp\left\{- \frac{(p+p')^2}{2\Delta p^2}\right\} \right].
\label{eq:transitionFinal}
\end{eqnarray}
The transition probability for negative $p$ can now be set to zero.
Moreover, at $p=0$,  a delta-function is placed, whose prefactor is chosen
such that the integral over the complete transition probability is unity. 

An interesting property of the transition probability approach is that
the distribution at any given magnification $\zeta$ can be calculated
from \eref{eq:transitionFinal}.
This is done by summing over
all  $\Delta p^2$ contributions coming from wavevectors with
$\vert {\bf q} \vert < \zeta q_{\rm l}$ into the transition matrix to yield
$\Delta p^2_{\rm tot}(\zeta)$. 
Moreover, the initial condition ($\zeta = 1$) for smooth interfaces is that the 
pressure is homogeneous across the interface when no roughness features are resolved.
Thus, it can be expressed as ${\rm Pr}(p,\zeta=1) = \delta(p-p_0)$, 
with $p_0 = L/A_0$, where $L$ is the normal load and $A_0$ the nominal contact area. 
This leads to 
\begin{eqnarray}
{\rm Pr}(p>0, \zeta)  = &
\frac{1}{\sqrt{2\pi\Delta p_{\rm tot}^2(\zeta)}} 
\left[ \exp\left\{- \frac{(p-p_0)^2}{2\Delta p^2_{\rm tot}(\zeta)}\right\} \right.
\nonumber\\ 
& \left.- \exp\left\{- \frac{(p+p_0)^2}{2\Delta p^2_{\rm tot}(\zeta)}\right\} \right].
\label{eq:pressureDistOri}
\end{eqnarray}
We use the variable $p$ (without subscript $0$)
for microscopic pressures while $p_0$ refers to the ``macroscopic'' pressure.

In the original version of the theory, the magnification dependent relative 
contact area $a_{\rm r}(p_0,\zeta)$ is obtained by integrating over the pressure 
distribution from infinitesimally small positive pressures to infinity, yielding
\begin{equation}
a_{\rm r}(p_0,\zeta) = {\rm erf}
\left\{ \frac{p_0}{\sqrt{2}\Delta p_{\rm tot}(\zeta)}\right\}.
\end{equation}

In more recent versions of the theory~\cite{Yang08JPCM}, Persson introduced a correction to
\eref{eq:broadening} using the argument that the broadening for
partial contact is less than for full contact. 
This leads to a modified broadening pressure
\begin{equation}
\Delta p^2_{\rm mod} = S\{a_{\rm r}(p_0,\zeta)\} \Delta p^2,
\end{equation}
where the ``fudge factor'' $S\{a_{\rm r}(p_0,\zeta)\}$ is parameterized
to match the numerical results for the pressure-dependence of the relative
contact area.
The following functional form has been used
\begin{eqnarray}
  S(p_0,\zeta)= [\gamma + (1-\gamma)a^2_{\mathrm{r}}(p_0,\zeta)],
\label{eq:gamma}
\end{eqnarray}
with $\gamma \approx 0.42$.
Thus, $S\{a_{\rm r}(p_0,\zeta)\}$ decreases monotonically from 
$S\{a_{\rm r}(p_0,\zeta)=1\}=1$ to $S\{a_{\rm r}(p_0,\zeta)\to 0^+\} \approx 0.42$.
This means that the modification is of order unity and thus relatively minor
given that many quantities span many decades, for instance the contact area, 
contact stiffness, and the spatial scales of the relevant wavelengths. 

With this modified broadening pressure, we cannot write down a closed-form 
expression for the {\it modified} version
of the {\it total} pressure broadening.
It now has to be determined self-consistently.
However, we may still use the formulae for the (total) pressure distribution
and the relative contact area, as long as we insert the corrected pressure
broadening terms. 

Up to this point, the uncontrolled approximations in Persson theory are:
(a) the pressure broadening $\Delta p^2$ (and similarly $\Delta p^2_{\rm mod}$)
entering the transition probabilities ${\rm Pr}(p,\zeta+\Delta\zeta\vert p',\zeta)$
only depend on $p-p'$ (and potentially on $a_{\rm r}$) but not on the 
initial pressure $p'$ of a walker in any other form,
(b) there is no drift in pressure at any value of $p'$, neither before nor after
adding the mirror Gaussian, and 
(c) there is no flow of the probability density at $p=0$ back to positive
pressure.
In the interpretation of the diffusion equation, it means that any walker gone 
out of contact is assumed to remain out of contact for good. 

Another quantity of interest is the elastic energy stored in the interface, 
${\cal E}$, which is needed, for example, in the derivation
of how the contact stiffness depends on pressure. 
According to the original Persson work~\cite{Persson01adhes}, the energy in a resolved mode is
\begin{eqnarray}
{\cal E}_{\rm P}(p_{0},{\bf q}) &= \frac{E^*}{4}\; q \; a_{\rm r}(p_0,q/q_{\rm l})\; \vert \tilde{h}({\bf q})\vert^2,\label{eq:elasticEnergy_h_q}
\end{eqnarray}
while unresolved modes are assumed to carry no energy.
In more recent work~\cite{Persson07PRL,Yang08JPCM}, the elastic energy was also modified with a correction factor to read
\begin{eqnarray}
{\cal E}_{\rm c}(p_{0},{\bf q}) &= {\cal E}_{\rm P}(p_{0},{\bf q})\,S\left\{a_{\rm r}(p_0,q/q_{\rm l})\right\}.
\end{eqnarray}
We denote the total energies stored in the interface by
\numparts
\begin{eqnarray}
{\cal E}_{\rm P}(p_{0}) &=  \sum_{\bf q} {\cal E}_{\rm P}(p_{0},{\bf q}), \label{eq:elasticEnergy_h_tot}\\
{\cal E}_{\rm c}(p_{0}) &=  \sum_{\bf q} {\cal E}_{\rm c}(p_{0},{\bf q})\label{eq:elasticEnergy_h_tot_wS}.
\end{eqnarray}
\endnumparts
Note that the contact area only depends on the nominal external load $p_{0}$
and the magnification $\zeta = |{\bf q}|/q_{\rm l}$.
Only the total sum \eref{eq:elasticEnergy_h_tot} or \eref{eq:elasticEnergy_h_tot_wS} needs to 
be accurate in the calculations relating to 
contact stiffness and mean gap, and not each individual term of \eref{eq:elasticEnergy_h_q}.
It might be necessary for other applications, such as rubber friction,
to impose stricter requirements.
Each summand associated with a given wave vector
should match, on average, the corresponding term of the exact elastic energy 
\numparts
\label{eq:elasticEnergy_u}
\begin{eqnarray}
{\cal E}_{\rm exa}(p_{0},{\bf q}, \zeta) &= 
\frac{E^*}{4}\; q\; \vert \tilde{u}(p_{0},{\bf q, \zeta}) \vert^2,
\label{eq:elasticEnergy_u_q}\\
{\cal E}_{\rm exa}(p_{0},\zeta) &= 
\sum_{\bf q} {\cal E}_{\rm exa}(p_{0},{\bf q}, \zeta),\label{eq:elasticEnergy_u_tot}
\end{eqnarray}
\endnumparts
where the $\tilde{u}(p_{0},{\bf q}, \zeta)$ are the exact elastic displacements in 
Fourier space for a given magnification and external load $p_{0}$. Those we determine 
to high accuracy from numerical simulations for
a given realization of surface roughness defined by the $\tilde{h}({\bf q})$. 

\subsection{Note on contact stiffness and Persson theory}

We wish to reemphasize that the purpose of this work is to analyze
the starting hypotheses of Persson theory rather than the final, 
experimentally measurable results arising from it.
It may yet be useful to remind the reader that some of these final results 
are the subject of current debate, in particular how the contact stiffness
$\kappa$ or the linearly related contact conductance depend on pressure.
Paggi and Barber~\cite{Paggi11} pointed out that many previous works
found a power law relation
\begin{equation}
\kappa \propto p^{\alpha}
\label{eq:scalingStiffness}
\end{equation}
with an exponent $\alpha < 1$ in agreement with their dimensional analysis
but in contradiction to a linear relation, $\alpha=1$.
The latter is yielded by the original Persson theory that implicitly assumes 
the thermodynamic limit, and also found in continuum simulations in which
self-affine roughness spreads only two decades~\cite{Campana11,Akarapu11}. 
Scaling arguments proposed by Pohrt {\it et al.}~\cite{Pohrt12pre}, which 
are meaningful when contact lives only in a single meso-scale asperity, lead
to an exponent $\alpha$ in \eref{eq:scalingStiffness} that solely
depends on the Hurst roughness exponent via 
\begin{equation}
\alpha = \frac{1}{1+H}.
\label{eq:stiffnessExponent}
\end{equation}
Their own numerical results, which were based on the so-called method
of dimensional reduction, could not confirm this result and instead
indicated that $\alpha \approx 0.266\,(3-H)$.
However, work based on accurate GFMD simulations as well as 
an extension of Persson theory to finite systems~\cite{Pastewka13}
found \eref{eq:stiffnessExponent} to be indeed true for pressures that
are so small that contact does not spread over the interface but is
located within a single meso-scale asperity. 
For more details, we refer the reader to the original 
literature~\cite{Paggi11,Pohrt12pre,Pastewka13,ProdanovDappMueser2014a}.

\subsection{Numerical methods}

We use Green's function molecular dynamics~\cite{Campana06,Kong09} (GFMD) to 
calculate the response of an ideally-elastic solid to deformations caused by 
mechanical contact with a rough counter body.
The solids are integrated over the $z$ coordinate and modeled as semi-infinite 
half-planes with hard-wall interactions in the small-slope approximation
so that only normal coordinates need to be considered. 
The setup is thereby reduced from a three-dimensional elasticity problem with 
$(3 {\cal L})^3$ independent variables to a classical boundary value problem with 
${\cal L}^2$ grid points, where ${\cal L}$ is the linear dimension of the system. 
We solve it with a molecular dynamics approach in reciprocal space, in order to 
reduce critical slowing-down from $\Or({\cal L}^2)$ to $\Or({\cal L}^{1/2})$. 
While a dynamic setup is possible, for this work we are interested mainly 
in the static limiting case and therefore use damped dynamics here.

The details of the method can be found in Ref.~\cite{ProdanovDappMueser2014a}.  
We use the parallel FFTW library~\cite{Frigo05fftw} which scales to several 
thousand cores. 
This allows us to tackle very large systems which is necessary to cover up to 
$5$ decades in roughness, close to what is found in natural or industrial 
surfaces~\cite{Power91,Persson05JPCM,Lechenault10}.
A simulation with linear size $2^{17}$ corresponds to $5\times 10^{15}$ 
(super-)atoms in an equivalent three-dimensional simulation.
More in-depth detail can be found in the original 
literature~\cite{Campana06,Kong09}.

\section{Results}\label{sec:results}

\subsection{Preliminary remark on relative contact area and the use of units}

\begin{figure}[htb]
\includegraphics[width=\textwidth]{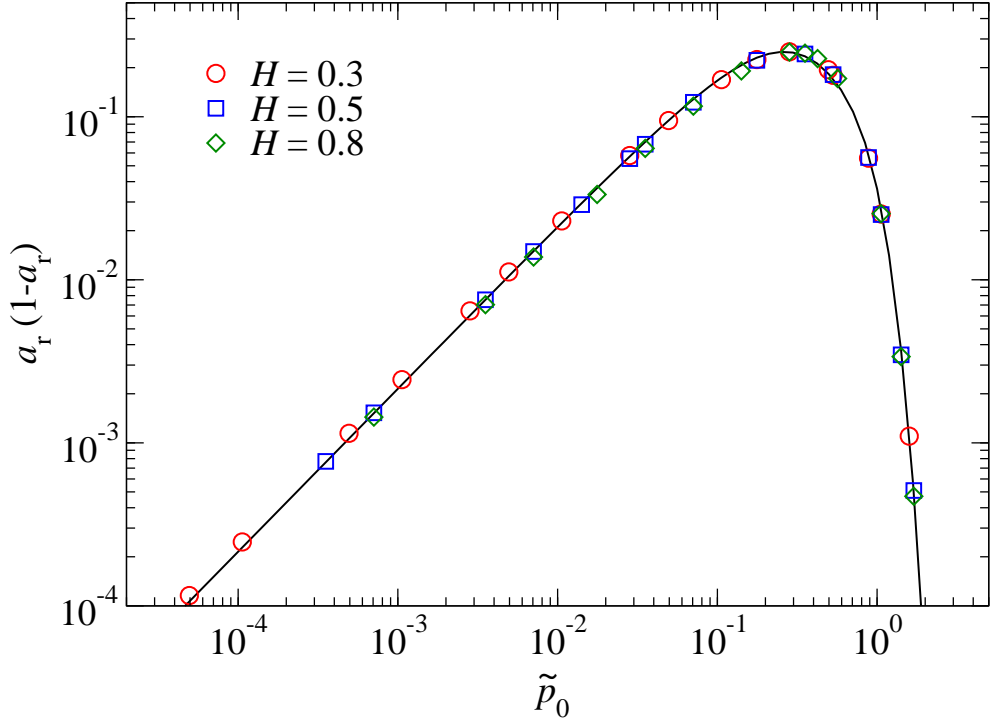}
\caption{ \label{fig:contactArea}
Representation of the relative contact area $a_{\rm r}$ as a function of the
reduced pressure for three different Hurst exponents $H$,
as obtained in previous GFMD simulations~\cite{ProdanovDappMueser2014a},
and the approximative function, \eref{eq:contactArea}.
To show the complementary relative contact area $(1-a_{\rm r})$ in the 
same figure, we plot $a_{\rm r}(1-a_{\rm r})$ rather than $a_{\rm r}$. 
}
\end{figure}

Before we present our tests of the assumptions made in Persson theory, 
we comment on the dependence of contact area on pressure and 
resolution, as well as on our choice for pressure.
In our previous work~\cite{ProdanovDappMueser2014a}, we found that
few dimensionless quantities suffice to define a contact.
In particular, we noticed that the contact area is essentially only a function
of reduced pressure $\tilde{p}_0 = p_0/E^*\bar{g}$ as long as the linear dimension
of the system ${\cal L}$ much exceeds the large-wavelength cutoff 
$\lambda_{\rm l}$ and the ratio of the cutoffs at small and
large wavelengths is sufficiently large. 
As demonstrated in \fref{fig:contactArea}, we can approximate the data 
from our previous work via the constitutive relation
\begin{equation}
a_{\rm r} \approx 
\{1-s(\tilde{p}_0)\}\; {\rm erf}(c_1\sqrt{\pi}\tilde{p}_0) +
s(\tilde{p}_0) 
\;{\rm erf}(c_2\sqrt{2}\tilde{p}_0),
\label{eq:contactArea}
\end{equation}
with  two fit parameters $c_1 = 1.075$ and $c_2 = 1.025$ being very close
to unity, and using a ``switching function''
\begin{equation}
s(\tilde{p}_0) = {\rm erf}^2(c_2\sqrt{2}\tilde{p}_0).
\label{eq:switching}
\end{equation}

The motivation for the functional form is that $a_{\rm r}$ can be described by a single error function
in the limit of low pressure and the complementary contact area by a single
(complementary) error function in the limit of large $\tilde{p}_0$. 
However, the numerical coefficients to be used in the error functions at small and
large pressure differ slightly. 
This is why we introduce a switching function that is close to zero at small
$\tilde{p}_0$ and close to one at large $\tilde{p}_0$ making the first summand
on the r.h.s. of \eref{eq:contactArea} dominate the sum for small $\tilde{p}_0$
and the term  ${\rm erf}(c_2\tilde{p}_0)$ be dominant at large $\tilde{p}_0$. 
(Since the leading correction to the linear low-pressure 
$a_{\rm r}\propto \tilde{p_0}$ relation is third-order in pressure,
we chose the switching function as the square of an error function.)
Previous simulations found that $a_{\rm r}\propto \kappa \tilde{p}_0$
with $\kappa \gtrsim 2$ for $\tilde{p}_0 \to 0$.
Our approximation for  $\kappa = 2.15 \equiv 2\, c_1 $ is close to that value.
The second term was written such that it describes the complementary contact
area at large pressures in exact accordance with Persson theory for $c_2 = 1$.
In agreement with a previous numerical study~\cite{YastrebovEtAl2014}, we find 
that a small correction needs to be applied. 
In principle, the coefficients $c_1$ and $c_2$ could be optimized for
different values of $H$, but for the stated numbers, $a_{\rm r}$ and
$1-a_{\rm r}$ are reproduced within $O(10\%)$ accuracy for any
value of $H = 0.3$, $0.5$, and $0.8$ used in the simulations. 

In the current work, we keep changing the resolution and thus it 
would not be meaningful to state absolute values of $p_0$.
It would not be meaningful either to express pressure as $p_0/E^*\bar{g}$,
because each time we increase the magnification at constant $p_0$,  
$\bar{g}$ would increase as well and thus the reduced pressure would change, 
although the absolute pressure $p_0$ would have remained unaltered. 
We therefore state or plot the relative contact area at a given magnification
rather than $p_0$ or $\tilde{p}_0$. 
With the help of \fref{fig:contactArea} or \eref{eq:contactArea}, 
these numbers can be easily converted into reduced pressures at that magnification.
Furthermore, in most cases one can simply associate $\tilde{p}_0 \approx E^*\bar{g} a_{\rm r}/2$.
We choose $E^*\bar{g}$ as unity to nondimensionalize the pressure~\cite{ProdanovDappMueser2014a}.

For microscopic, local pressures, i.e., those that hold for individual points
at the interface, we use $p_0 / a_{\rm r}$ as default unit as the latter reflects
the mean pressure averaged over the contact.
Thus, when identifying a walker with a pressure much less than unity,
there is a large probability that it sits
either close to a contact line or in a small patch bearing little load. 
As stated before, the variable $p$ stands for local pressures 
while $p_0$ refers to the ``macroscopic'' pressure.

\subsection{Pressure-independent and drift-free broadening}
\label{sec:driftDiffusion}

In this section we test the first two approximations implicitly contained 
in Persson theory.
They can be described as the following two properties of 
the transition probability ${\rm Pr}(p',\zeta+\Delta\zeta\vert p,\zeta)$:
(a) the term related to the broadening of the pressure distribution,
$\Delta p^2$ depends only on $p-p'$, i.e., the ``diffusion coefficient''
\begin{equation}
D = \Delta p^2/\Delta \zeta 
\end{equation}
is independent of $p'$,  and
(b) the transition probability induces no drift at any pressure. 

\begin{figure}[htb]
\includegraphics[width=\textwidth]{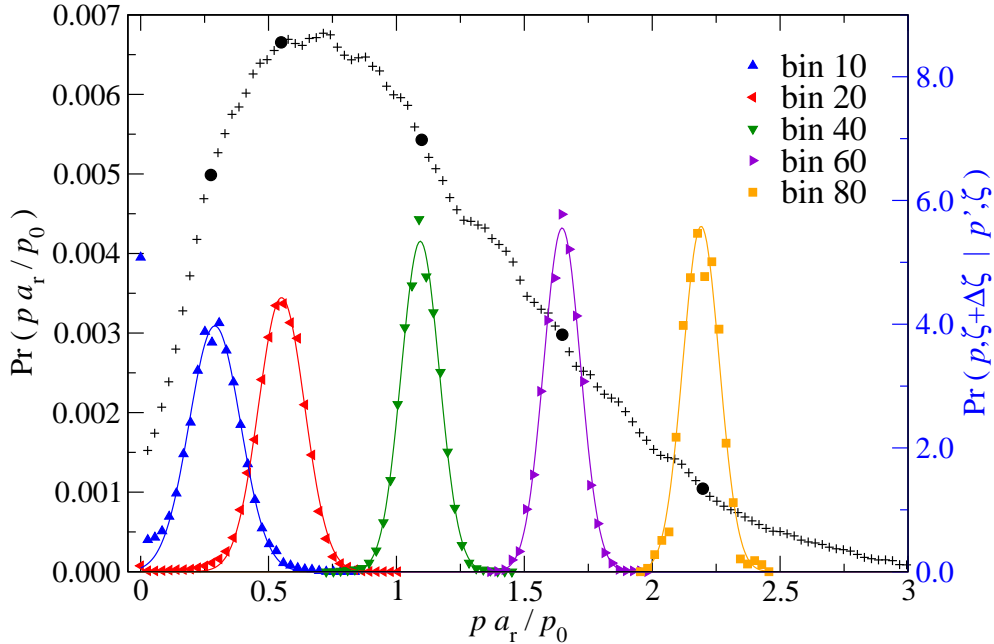}
\caption{ \label{fig:broadening}
Original, discretized pressure distribution ${\rm Pr}(p)$, indicated
by plus signs, at a magnification of $\zeta = 63$.
For selected bins, indicated by full circles, the transition probability 
${\rm Pr}(p,\zeta+\Delta\zeta\vert p',\zeta)$ is recorded for $\Delta \zeta = 0.3$ 
(right ordinate axis). They are shown in color.
Full lines represent fits to mirror Gaussians as described in the Appendix. Starting at bin 20, the 
difference to simple Gaussian is negligible. 
The heights and thus the widths of the transition probabilities
turn out to depend on pressure $p$. The simulations were run with $H = 0.5$ and $a_{\rm r} \approx 0.01$.
}
\end{figure}

In order to test these two approximations,
we first ran simulations with a maximum resolved target wave number
$\lambda_{\rm t} =  {\lambda_{\rm l}/\zeta}$ with $\zeta = 63$.
In these calculations, the Hurst exponent was set to $H = 0.5$ and
a system size of ${\cal L} = 16,384$ was investigated. 
The pressure was chosen such that it produced a relative contact area of $0.01$
for the given magnification of $\zeta = 63$. 
From the relaxed configurations we computed the pressure at each point
at the interface and produced a pressure distribution function from it
on a discretized mesh with constant spacing $\Delta p$. 
For selected bins with index $n$, we memorized each grid point, or walker, 
for which the pressure lay in the interval $n\Delta p \le p < (n+1) \Delta p$ 
at $\zeta = 63$.
The magnification was then increased to $\zeta = 63.3$ and the 
distribution function (of the new configuration) evaluated over the points that had been associated
with the given bins at the old magnification.
This yields a discretized version of the transition probability.
Results are shown in \fref{fig:broadening}. 

\begin{figure}[htb]
\includegraphics[width=\textwidth]{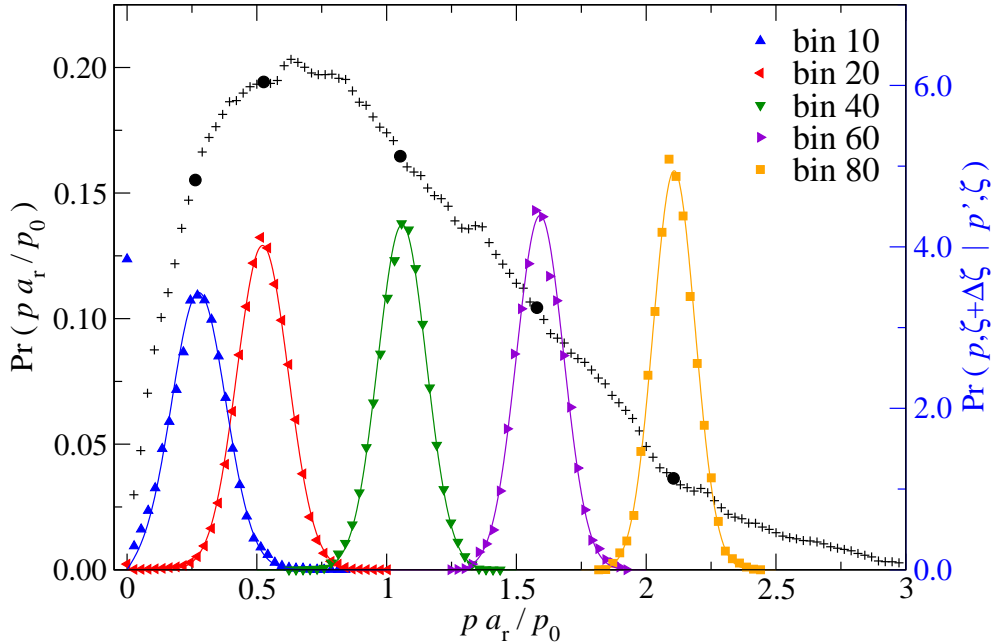}
\caption{ \label{fig:broadening2}
Similar to \fref{fig:broadening}, except that the simulations used here were
for $H = 0.8$ and $a_{\rm r} \approx 0.31$, and the magnification changed from $\zeta=16$ to $\zeta=16.2$. 
The figures show identical trends.
}
\end{figure}

\begin{figure}[htb]
\includegraphics[width=\textwidth]{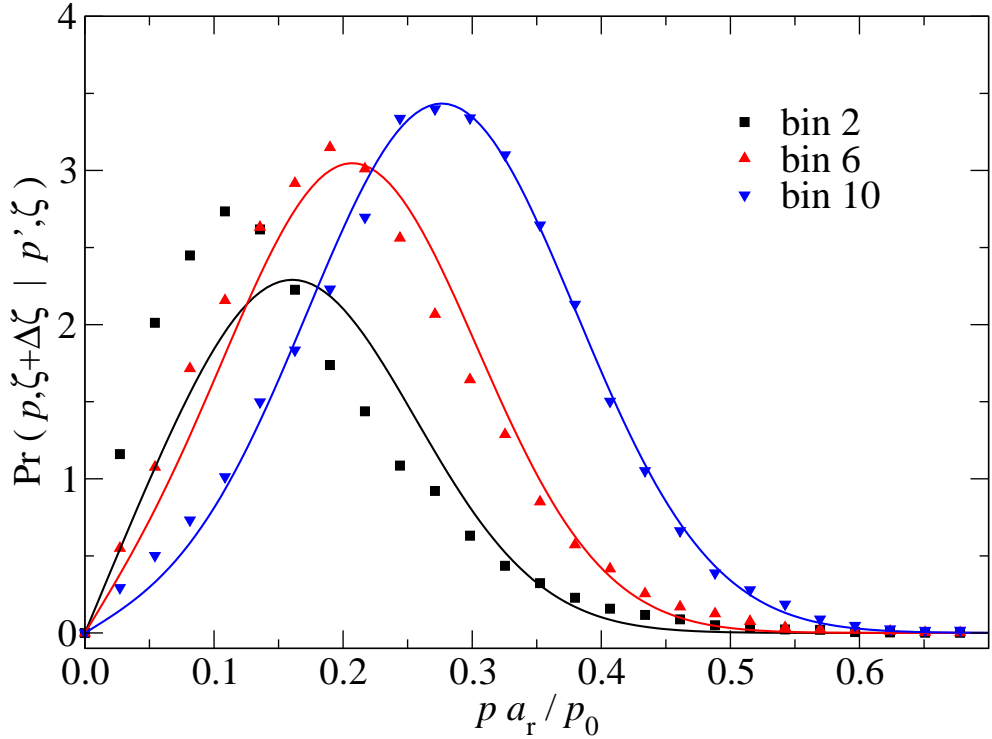}
\caption{
Pressure distribution of some low-pressure bins for $H = 0.8$ and $a_{\rm r} \approx 0.31$. 
The singularity at $p = 0$ is omitted in all cases.
The mirror-Gaussian function with on-the-fly determined parameters (solid lines) only fits well as long
as the broadening is smaller than the mean. Below that threshold, the description is unsuitable. 
It is in principle still possible to fit a mirror-Gaussian to the data (not shown), but the 
normalization is inconsistent. We therefore did not include bins $<8$ in our analysis.
\label{fig:histogram_lowPbins}
}
\end{figure}

In \fref{fig:broadening}, the transition probabilities are described
accurately by mirror Gaussians, for the bins whose mean pressure 
much exceeds the broadening, and information on the diffusion coefficient 
can be ascertained directly. 
For example, a systematic on-the-fly determination of the diffusion
coefficient associated with one of these bins can be done by subtracting the
variance obtained at the old magnification from that at the new magnification.
However, special care has to be taken for the analysis of those bins
representing small pressures. 
In \ref{sec:onTheFly} we describe how to compute drift and diffusion 
coefficients such that their determination is also meaningful when the
mean pressure of a bin is smaller than the broadening.
If the diffusion into the singularity at $p=0$ is negligible, even a simple Gaussian
suffices with high accuracy.

We carried out the same analysis for $H = 0.8$ and a much larger relative
contact area of $a_{\rm r} \approx 0.31$. \Fref{fig:broadening2} shows
that there is no qualitative difference for the different Hurst exponents, or at 
different external loads.

\Fref{fig:histogram_lowPbins} shows bins for which the broadening is larger
than the means. In this case, the procedure laid out in \ref{sec:onTheFly} 
is \textit{not} suited anymore. While bin 10 is still described quite well
(but not by a simple Gaussian anymore), for bin 6, only the width is still acceptable. 
The peak is shifted slightly. For bin 2, finally, neither the width nor peak are suitably 
described. We did not include any data from bins $< 8$ in the following.

\begin{figure}[htbp]
\includegraphics[width=\textwidth]{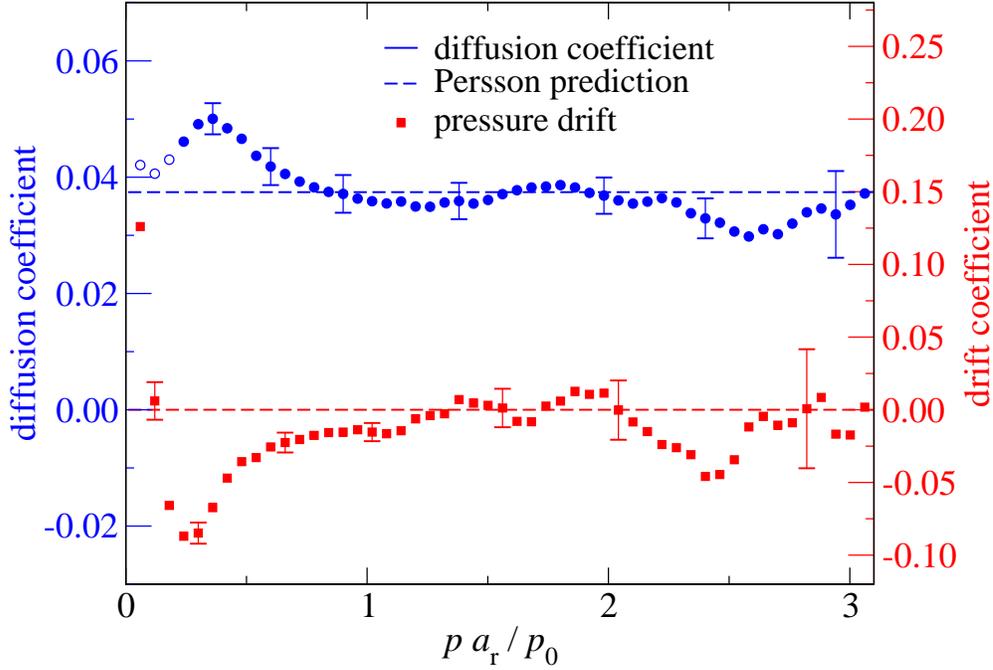}
\caption{ \label{fig:broadening_drift_pressure}
Diffusion (blue circles, left $y$-axis) and drift coefficients (red square, right $y$-axis),
describing the evolution of the pressure distribution between 
a magnification of $16$ and $16.2$, at relative contact areas 
of $a_{\rm r}(\zeta=16) = 0.2667$ and $a_{\rm r}(\zeta=16.2) = 0.2662$.
%
The shown data is averaged over four equivalent but statistically independent 
realizations of the surface.
The dashed lines reflect assumptions made in the original version of
Persson theory. 
Non-negligible stochastic uncertainties remain for $p \gg p_0/a_{\rm r}$ as 
revealed by the error bars.
Stochastically significant discrepancies from the drift-free and
the constant-diffusion-coefficient assumptions remain for
pressures $p \ll p_0/a_{\rm r}$. 
These can be rationalized by assuming that low-pressure points tend to
be located close to a contact line. 
Open symbols are from bins where the mirror-Gaussian fit function is unsuited
to determine the diffusion coefficient, even though the drift 
in those bins is well defined.
}
\end{figure}

In order to arrive at more quantitative results, we conducted a moments analysis 
of the pressure distribution, as described in \ref{sec:onTheFly},
for $64$ bins, {for a relative contact area of $a_{\rm r} = 0.3$}. 
The results shown in \fref{fig:broadening_drift_pressure} reveal that the
drift is negligible for large pressures and that the diffusion coefficient
comes out as assumed in the \textit{original} version of Persson theory.
That means that
no correction factor is required to predict the correct broadening for most
values of $p$,
although for $a_{\rm r} = 0.3$, $S(p_0,\zeta)$ in \eref{eq:gamma} should
already be close to its minimum value $\approx 0.42$.
Drift and diffusion coefficients only deviate from the prediction
for $p \ll p_0/a_{\rm r}$, that is, at pressures much smaller than 
the mean pressure in the contact regions.

It can be easily rationalized why the assumptions made in Persson theory
hold for $p \gtrsim p_0/a_{\rm r}$ but not for $p \ll p_0/a_{\rm r}$.
Walkers contributing to the histogram at pressures $p \gtrsim p_0/a_{\rm r}$ 
lie far away from any contact line and pressure gradients should usually be small.
The assumption that additional roughness leads to small pressure perturbation
is thus justified, certainly as long as $\lambda_{\rm l}/\zeta$ is small
compared to the linear dimension of the contact patch to which this walker belongs.  
In contrast, walkers contributing to the histogram at $p \ll p_0/a_{\rm r}$
lie close to a contact line, or, more generally, close to a point or patch 
that risks to fall out of contact soon.
Pressure gradients are high at those positions and even diverge right at the
contact line (as in Hertzian contacts), which explains why the
diffusion coefficient picks up at small pressures. 
Moreover, pressure gradients increase as the contact line is approached, 
which is consistent with the presence of a negative drift.
If, however, a walker is {\it extremely} close  to a contact line, there
is a large probability that the walker jumps from contact with large
pressure gradients to out-of-contact, where the pressure gradient is zero.
This explains why the drift turns around in sign at extremely small pressures.

An interesting observation in \fref{fig:broadening_drift_pressure}
is that the pressure broadening in the contact agrees with the original,
correction-parameter-free
variant of Persson theory rather than with the modified version.
At the same time, points fall out of contact more quickly than predicted
by Persson theory because many walkers acquire a negative drift 
at $p < p_0/a_{\rm r}$.
Their number is distinctly larger than of those having a positive drift
so that the average {\it contact} drift coefficient
\begin{equation}
\bar{\mu}_{\rm c}  = \frac{1}{a_{\rm r}}\int_{0^+}^\infty {\rm d}p\, {\rm Pr}(p) \mu(p)
\end{equation}
is negative.

So far, our calculations imply that there is mean drift towards smaller pressures
and an increased diffusion at small pressure as compared to the theoretical
prediction.
From this point of view, one would expect Persson theory to overestimate the
contact area. 
However, the opposite is true.
Thus, one must expect walker to re-enter contact upon an increase of magnification,
which would counteract the large flux out of contact that is induced
by large diffusion coefficients and negative drifts at small $p$.
We investigate this hypothesis next.

\subsection{No re-entry}

\begin{figure}[htb]
\includegraphics[width=\textwidth]{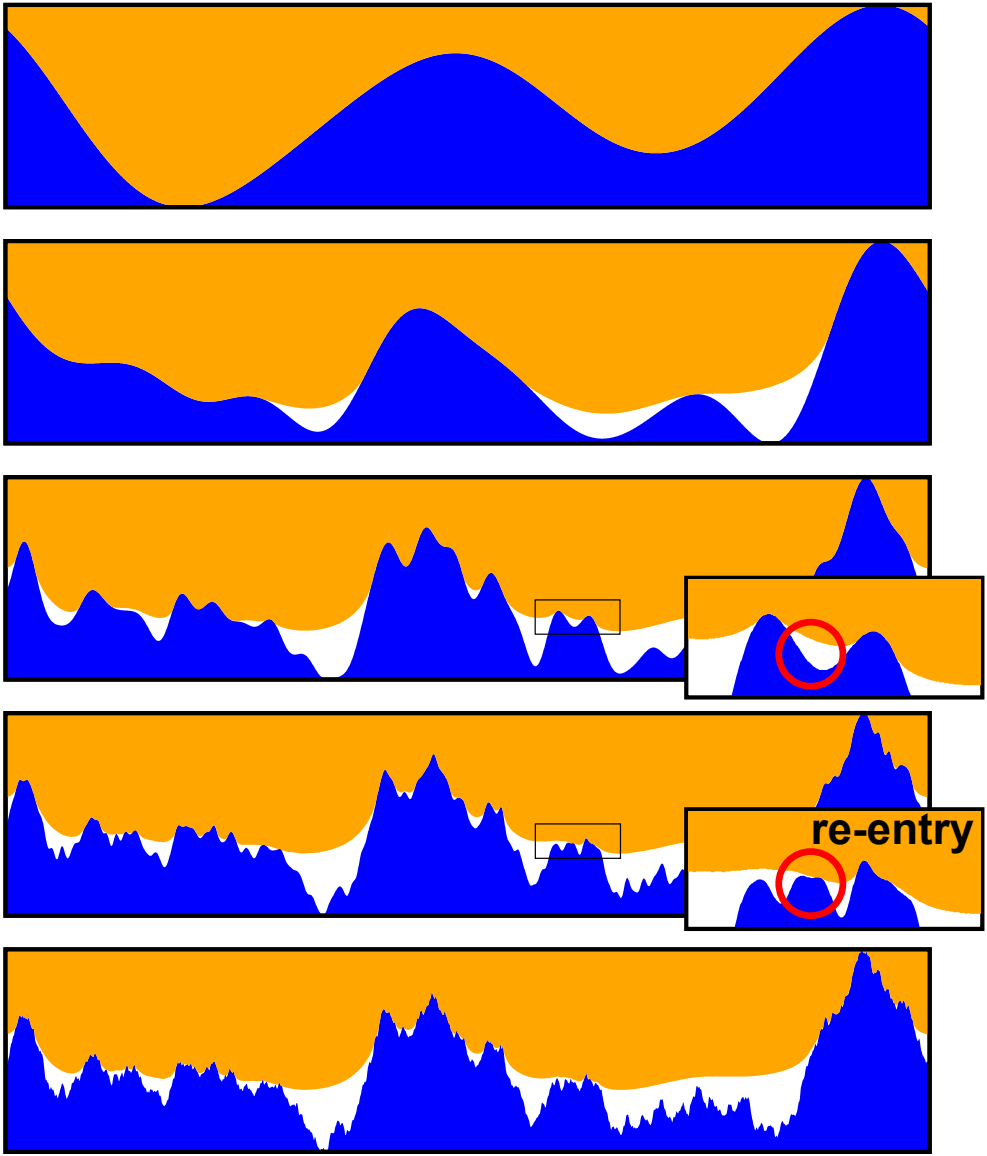}
\caption{ \label{fig:reentry_waterfall}
(Color online)
Cut through a rough contact with gradually increased roughness on
smaller and smaller scales. 
The magnification increases from top to bottom with $\zeta = \{2, 4, 16, 64, 256\}$.
The top (orange) solid holds all the compliance, 
while the lower (blue) body contains all the roughness. 
The gap is kept in white. 
Even though the overall shape does not change at large magnification, 
the local topology does, and 
this causes re-entry of some points that had previously fallen out of contact.
}
\end{figure}

One key assumptions in Persson theory is that points at the interface 
are seen to fall out of contact as finer details of the surface roughness
are incorporated. 
The idea is visualized in \fref{fig:reentry_waterfall}, which shows
cuts through the converged surfaces for simulations with system sizes of 
$4,096 \times 4,096$ at five different 
resolutions $\zeta = \{2, 4, 16, 64, 256\}$.
The cuts are at the same location for each surface. 
In each panel, roughness is added on successively smaller scales. 
Even though we did not select the location of the cut specifically for this 
effect, we happened to find a re-entry point. 

\begin{figure}[htb]
\includegraphics[width=\textwidth]{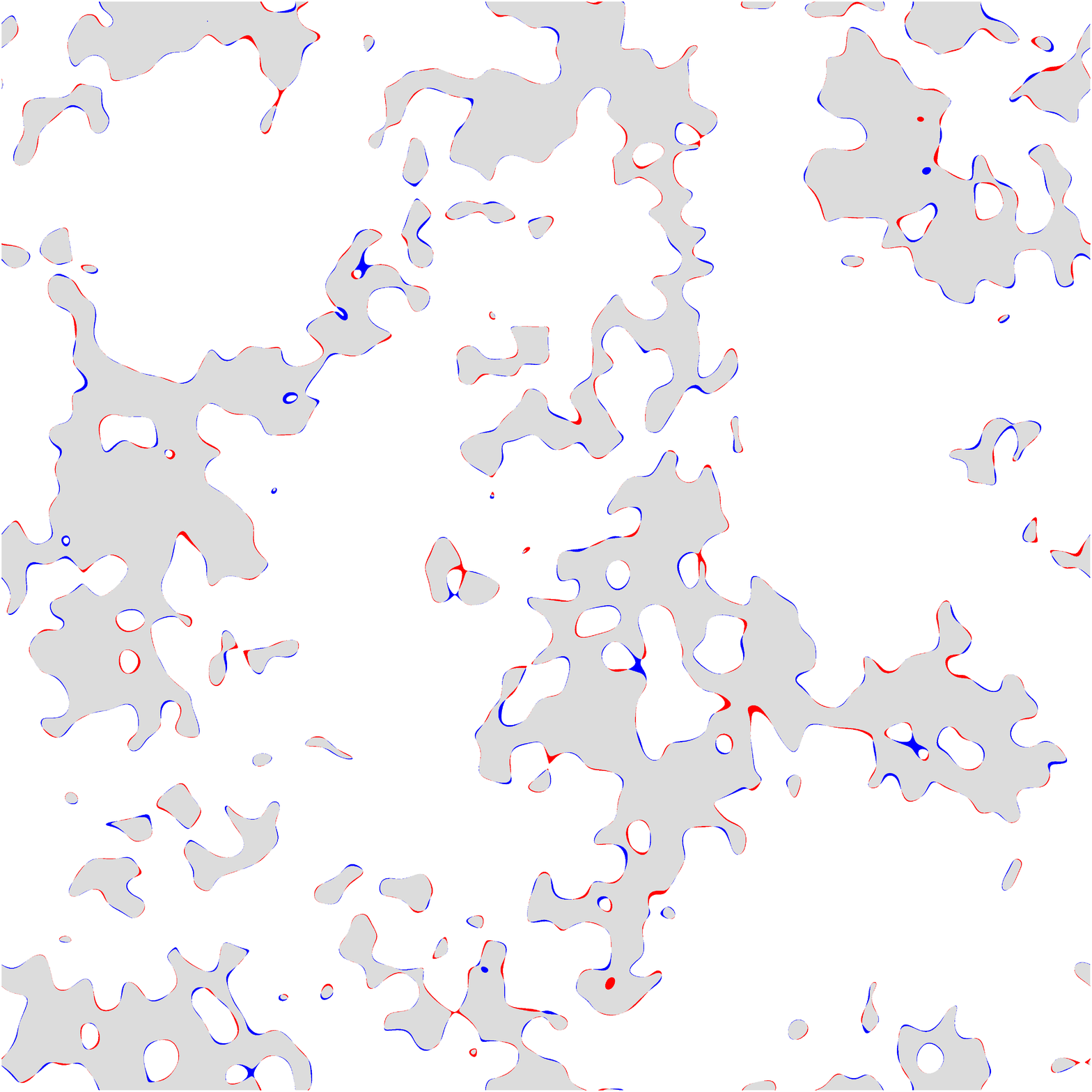}
\caption{ \label{fig:reentry_visualization}
(Color online)
Visualization of a contact geometry of a  $4,096 \times 4,096$ system with
$H=0.8$ and a relative contact area of $a_{\rm r} = 0.14$ at 
magnification $\zeta = 129.5$.
White and blue areas represent non-contact, while gray and red show contact patches.
Blue and red points have changed their nature between 
$\zeta = 128$ and $\zeta = 129.5$, i.e., blue was in contact at $\zeta = 128$,
while red was in non-contact at the smaller magnification. 
Since all points are in contact when only the very longest wavelength is 
considered ($\zeta = 1$), the red points have ``\textit{reentered}'' contact. 
The relative contribution of red and blue are $0.675~\%$ and $0.597~\%$ of the
apparent contact area, respectively,
implying that the net flow $0.078~\%$ is small compared to the flux in 
either direction. 
The reentry process is not accounted for by Persson theory. 
}
\end{figure}

A similar analysis as that for the cross section  of a contact shown in
\fref{fig:reentry_waterfall} is repeated in terms of a bird's eye view of 
the interface in \fref{fig:reentry_visualization}.
There we highlight areas that change from contact to non-contact and vice versa, 
as $\zeta$ is increased from $128$ to $129.5$.
Analysis of the data reveals that the net flux to non-contact is a small 
fraction of $\Or(15\%)$ of the flux in either direction.
The large flux of points getting back into contact supports the hypothesis
that reentry distinctly increases the contact area relative to the predictions 
by Persson theory.
The effect is sufficiently large to overcompensate the effects discussed
in Section~\ref{sec:driftDiffusion}.

As one may expect a re-entry contact point to get out of contact
soon again, it is not clear how many contact points at a given magnification
are reentrant points. 
To answer the question how many points have \textit{ever} returned to
contact --- or returned to non-contact --- one needs to examine
every single change in magnification.
This is an extremely tedious and computationally expensive procedure 
--- even for a small system, $1,024 \times 1,024$ in size, 
with $\lambda_{\rm l} = 512$ and 
$\lambda_{\mathrm{s}} = 4$, there are a total of $\approx 4,400$ different wave numbers, 
each of which has to be added one-by-one, and the same number of separate simulations run, 
for each value of the external load. 
In addition, this process has to be repeated for different surfaces to ensure 
robust numbers.
We ran a set of eight different random realizations
for each pressure value of size $1,024 \times 1,024$.
We did not attempt to carry out this analysis for systems larger than $4,096\times4,096$, 
so that the resulting percentage is still only an estimate, as the fractal limit is 
not reached fully.
However, it does serve as an orientation --- when taking into account
a larger range of magnifications more points are bound to experience re-entry
than for a smaller range. 
Nevertheless, for $a_{\rm r} = 0.14$ and $H = 0.8$, 
we find that $\approx 62\%$ of all contact points at the highest resolution 
have left and re-entered 
contact at a lower magnification. 
For $a_{\rm r} = 0.014$, \textit{every single} point in 
contact had lost contact at a lower magnification. 
Even for $a_{\rm r} = 0.88$, where contact is
nearly complete, the fraction is still $\approx 8\%$ and thus non-negligible.
These numbers do not change if the continuum limit is approached even closer ---
for $\epsilon_{\rm c} = 1/8$ and $1/16$, the fractions remain the same. 
They do not depend much on $\lambda_{\rm l}/{\cal L}$ either, and remain 
comparable for $\epsilon_{\rm t} = 1/4$.
The Hurst exponent similarly has a very minor effect; the numbers for $H = 0.3$
lay within $4\%$ of those for $H = 0.8$, and neither showed a dependence 
on $\epsilon_{\rm t}$ nor $\epsilon_{\rm c}$.

\begin{figure}[htb]
\includegraphics[width=\textwidth]{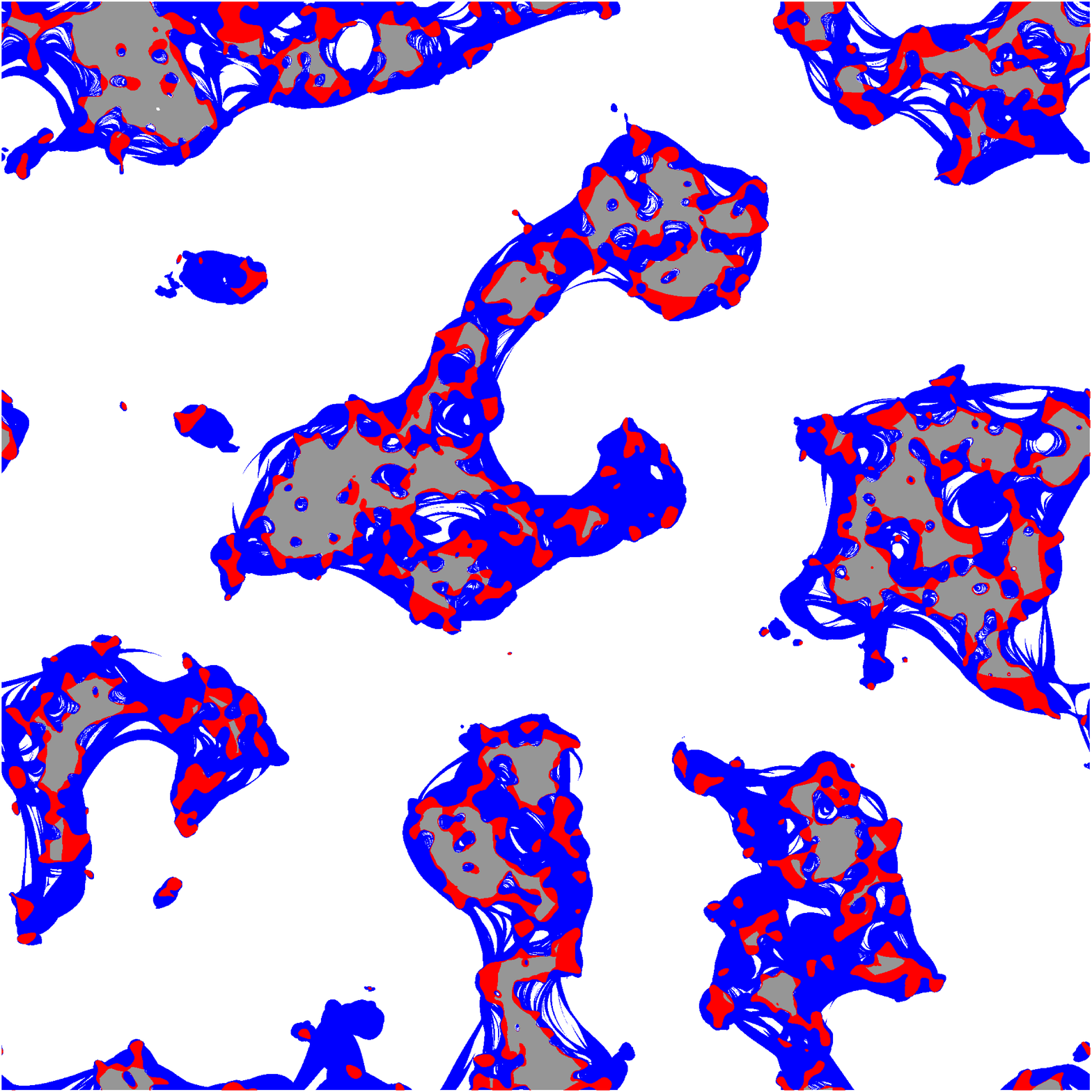}
\caption{
Similar to \fref{fig:reentry_visualization}, except for a $2048 \times 2048$ 
system and different initial ($\zeta = 3.2$) and final ($\zeta = 32$)
magnifications, for which the contact areas were
$a_{\rm r} = 0.28$ and 
$a_{\rm r} = 0.14$, respectively.  
Here, gray points have been in contact at all intermediate magnifications, 
and white points have either been out of contact at $\zeta = 3.2$ 
or lost contact exactly once.
In contrast, all points in color have experienced reentry between 
$\zeta = 3.2$ and $\zeta = 32$; 
blue points show out-of-contact reentry and red indicates contact reentry.
Note that there are $3.8$ times more colored points than gray points, and of all 
points in contact at the final magnification, $50\%$ have experienced reentry. 
Sharp features all come from low magnifications where the 
surface changes relatively strongly with each added wave number.
\label{fig:reentry_visualization_allQ}
}
\end{figure}

\Fref{fig:reentry_visualization_allQ} visualizes the reentry for 
a system of $2048\times2048$ over a larger range of magnifications. All points
shown in color have reentered contact in the examined range, and only 
$\approx 25\%$ of all points have remained in contact.

\begin{figure}[htb]
\includegraphics[width=\textwidth]{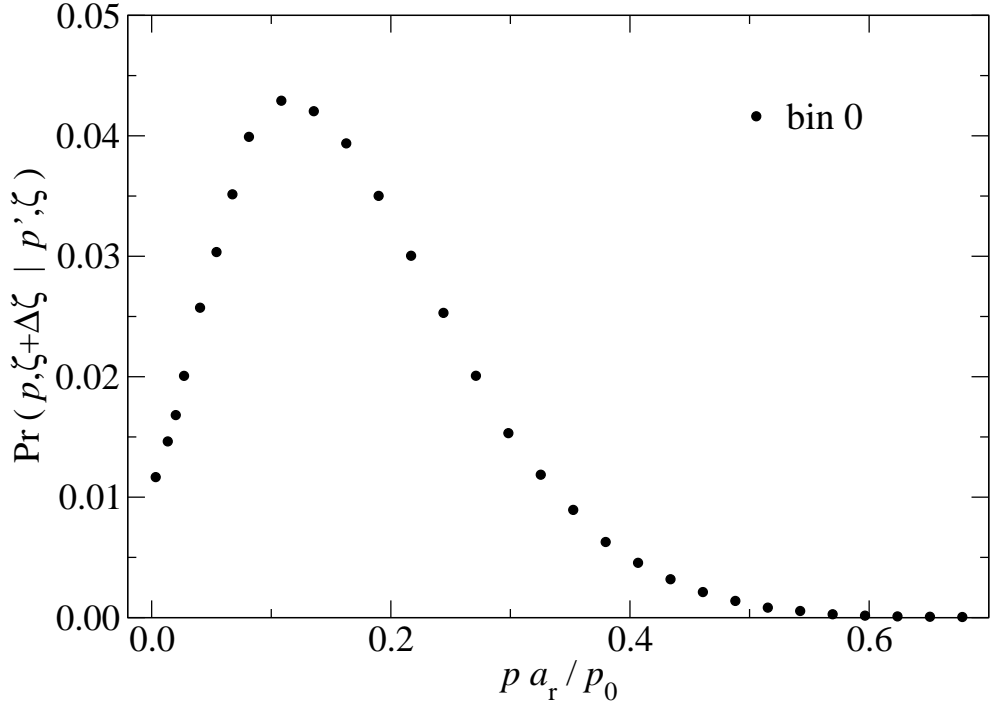}
\caption{
Histogram of the points in the zero pressure bin at $\zeta = 63$, after the magnification 
is increased to $\zeta = 63.3$. The total transition probability is $\approx 0.01$. 
This means that $1\%$ of all points reenter between $\zeta = 63$ and $\zeta = 63.3$.
\label{fig:reentry_histogram}
}
\end{figure}

To complete the analysis of reentrance, we note that \fref{fig:reentry_histogram} 
includes data for the pressure transition probability for an initial pressure of $p'=0$,
similar to our analysis for finite $p'$.
For a change from $\zeta = 63$ to $\zeta = 63.3$, we find about $1\%$ of all points
previously in noncontact reenter contact. 
Such points therefore act as if they came from  a ``source'' in the framework
of the diffusion analogy, or, more precisely, as if they were reflected 
--- potentially with a delay --- by the boundary. 
This can affect the functional form of the pressure distribution function
and lead to deviations from a linear ${\rm Pr}(p) \propto p$ dependence 
at small $p$, which one gets for the distribution shown in
\eref{eq:pressureDistOri}.
In fact, preliminary results are in violation of a ${\rm Pr}(p) \propto p$
relationship.

\subsection{Single-mode analysis of the elastic energy}

\begin{figure}[!p]
\includegraphics[height=0.9\textheight]{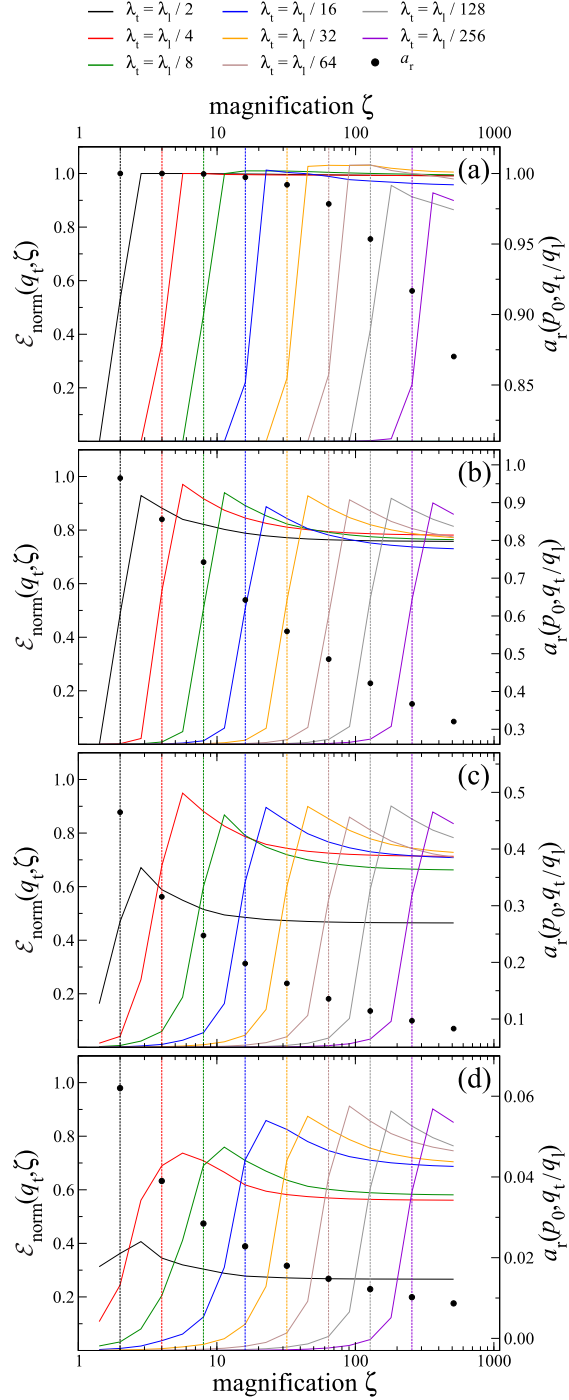}
\caption{
Normalized elastic energy, see \eref{eq:elastic_energy_norm} for system 
sizes of $16,384\times 16,384$ and a Hurst exponent of $H = 0.8$, averaged 
over 8 statistically equivalent but independent random realizations. The cutoff 
wave numbers were $\lambda_{\rm l} = 8,192$ and $\lambda_{\rm s} = 8$, 
approximately fulfilling the thermodynamic, fractal, and continuum limits. 
Each panel shows the elastic energy for a different relative contact area 
at full resolution (top to bottom) of 0.87, 0.32, 0.083, 0.009, in a collection 
of nearby Fourier modes with 
$q_{\rm t} \approx \zeta q_{\rm l}$ normalized by the energy that they are expected 
to have in the original Persson theory, i.e., normalized by 
$a_{\rm r}(q_{\rm t}/q_{\rm l})\;E^{*}/4 \sum_{\vert{\bf q}\vert\approx q_{\rm t}} 
q \vert \tilde{h}({\bf q}) \vert^2$.
The right ordinate axis shows the relative contact area at the given 
magnifications. 
\label{fig:elastic_energy_H08}
}
\end{figure}

\begin{figure}[!p]
\includegraphics[height=0.9\textheight]{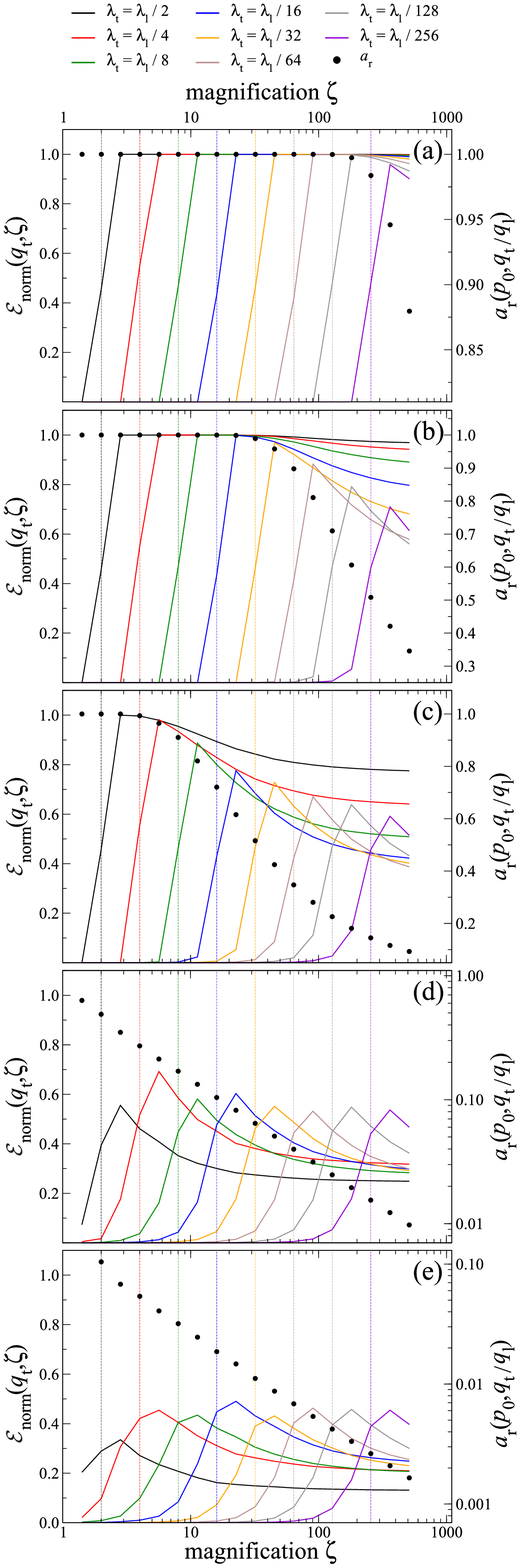}
\caption{
Similar to \fref{fig:elastic_energy_H08} but for $H = 0.3$ and the 
contact areas (top to bottom): 0.88, 0.35, 0.093, 0.0098, 0.0017. Note 
that in panels (d) and (e), the axis for the contact area is logarithmic.
\label{fig:elastic_energy_H03}
}
\end{figure}

As mentioned before, the derivation of Persson theory assumes that an energy mode contains 
no energy until the appropriate mode of roughness is resolved. Beyond this magnification, the
energy in this mode is expected remain the same when even greater wave numbers are included. 
For various external loads, 
\fref{fig:elastic_energy_H08} and~\ref{fig:elastic_energy_H03}
demonstrate that this is an oversimplification when 
contact is incomplete. 
We plot the following expression for a given target wave number $q_{\rm t}$
\begin{eqnarray} 
{\cal E}_{\rm norm}(p_{0},q_{\rm t}, \zeta) 
&\equiv \frac{{\cal E}_{\rm exa}(p_{0},q_{\rm t}, \zeta) }{{\cal E}_{\rm P}(p_0,q_{\rm t}) }\nonumber\\
&\approx \frac{\sum_{|{\bf q}|\approx q_{\rm t}} q\; \vert \tilde{u}({p_0,\bf q, \zeta}) \vert^2}
{\sum_{|{\bf q}|\approx q_{\rm t}}  q \; a_{\rm r}(p_0,q_{\rm t}/q_{\rm l})\; 
\vert \tilde{h}({\bf q})\vert^2},\label{eq:elastic_energy_norm}
\end{eqnarray}
while 
\begin{eqnarray} 
{\cal E}^{\rm pred}_{\rm norm}(p_{0},q_{\rm t}, \zeta) 
&= S(p_0,\zeta)\; \Theta(\zeta - q_{\rm t}/q_{\rm l})
\label{eq:elastic_energy_norm_predicted}
\end{eqnarray}
is the \textit{predicted} normalized energy. $\Theta(\zeta - q_{\rm t}/q_{\rm l})$ 
is the Heaviside step function which is zero for $\zeta < q_{\rm t}/q_{\rm l}$ and
$1$ for $\zeta \ge q_{\rm t}/q_{\rm l}$.
We find that each mode is partially excited already at a lower 
magnification than expected, and peaks at a higher magnification. 
Partly this is 
caused by averaging the contribution of a number of nearby wave numbers for reasons of 
reducing scatter. 
Another important effect should be related to the following phenomenon:
The derivative of the stress becomes singular as a contact line is approached
from within the contact.
It then discontinuously drops to zero outside the contact,
as one can readily see in the case of a Hertzian contact.
This implies that the Fourier coefficients of the stress and thus
the strain field must be non-zero up to (infinitely) large wavevectors 
as soon as there is partial contact. 
The corresponding amplitudes may be small, but they are non-zero. 

When the external load is very high, as is the case in 
\fref{fig:elastic_energy_H08}a, contact remains complete up to 
high magnification, and \eref{eq:elastic_energy_norm_predicted}
is essentially accurate.
When the pressure is reduced, as in \fref{fig:elastic_energy_H08}b,
or, alternatively, the magnification is increased further, 
less energy starts to be stored in the short-wavelength modes than predicted.
However, also the long-wavelength modes are populated less than
anticipated. 
Interestingly, the energy reduction is even stronger 
for long wavelengths than for short wavelengths and in contradiction
to the functional form of $S(p_{0},\zeta)$ in \eref{eq:gamma}, as
revealed by \fref{fig:elastic_energy_H08}c
and~\ref{fig:elastic_energy_H08}d.

Short wavelengths are populated in a way that roughly conforms with
Persson theory before including the correction factor \eref{eq:gamma}.
However, long wavelength displacements do not appear to develop
as much as expected and even appear to recede when roughness is 
added at large magnification and small external pressure. 
This effect is captured neither in the original version of Persson theory
nor in the modified version.

\begin{figure}[htbp]
\includegraphics[width=\textwidth]{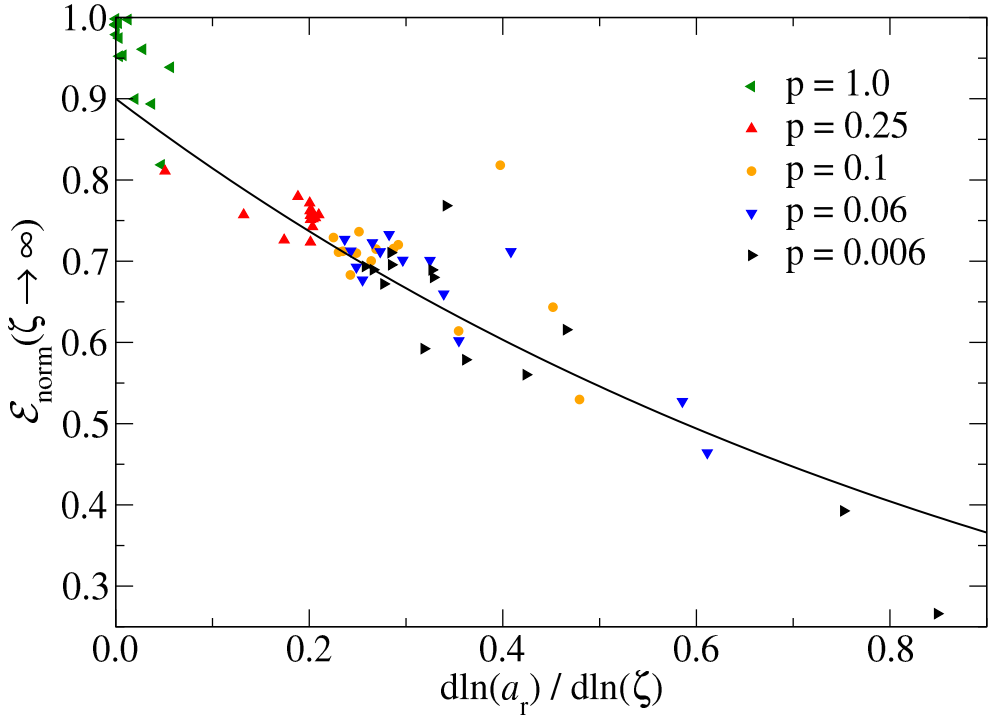}
\caption{ 
Asymptotic value of the ratio of elastic energy to Persson prediction for $H = 0.8$
(see \fref{fig:elastic_energy_H08}) versus the relative change of the relative
contact area with magnification, ${\rm d}\ln a_{\rm r} / {\rm d}\ln \zeta$. Despite
some scatter, a correlation is visible, of the form $0.9\,\exp(-1.0\,x)$.
}
\label{fig:asymptoticValue_vs_dlnArel_H08}
\end{figure}

\begin{figure}[htbp]
\includegraphics[width=\textwidth]{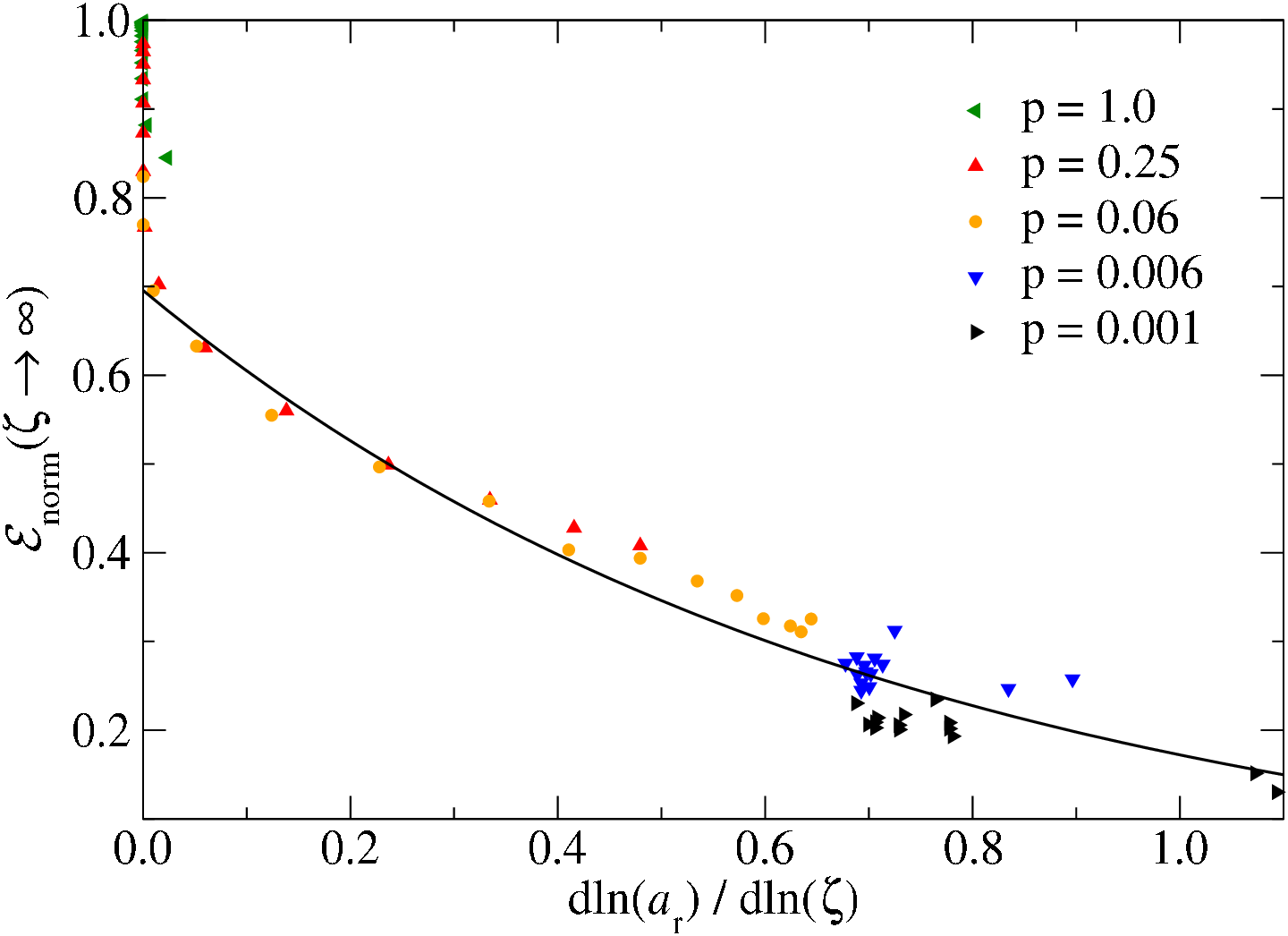}
\caption{ 
As \fref{fig:asymptoticValue_vs_dlnArel_H08}, but for $H = 0.3$. The correlation
is not the same as for $H = 0.8$, but the same form, with $0.7\,\exp(-1.4\,x)$.
}
\label{fig:asymptoticValue_vs_dlnArel_H03}
\end{figure}

\Fref{fig:elastic_energy_H08} and~\ref{fig:elastic_energy_H03} seem to 
show a correlation between the asymptotic value that the elastic energy, normalized
with Persson's prediction, converges to, and the relative change of the relative
contact area. We explore this correlation in \fref{fig:asymptoticValue_vs_dlnArel_H08}
and~\ref{fig:asymptoticValue_vs_dlnArel_H03}. The fall-off of the elastic energy
turns out to be linear with magnification, so we can extrapolate the curves
for $\lambda_{\rm t} \ge \lambda_{\rm l}/128$ to $\zeta \longrightarrow \infty$
to determine the asymptotic value.

Ignoring the values at full contact, there indeed is a correlation, despite some scatter, of the form
$\tilde{K}_{1}\,\exp \left[-\tilde{K}_{2}\,{\rm d}\ln (a_{\rm r}) / {\rm d}\ln (\zeta)\right]$.
The value of the constants is not universal and changes with Hurst exponent, but 
--- at least for the two cases we inspected --- assumes values of $\Or(1)$. 
Further investigations are necessary.

\subsection{Analysis of the integrated elastic energy}

The previous section showed that the individual modes of the elastic energy
do not behave as Persson theory assumes, except near perfect contact. Nevertheless
the theory quite accurately predicts the relative contact area and
the mean gap found in the contact between two randomly rough elastic bodies.
Especially the latter depends intimately on the elastic energy --- albeit not on 
each mode but on the total sum. Still it is not obvious that the total elastic
energy is correct while each individual term is inaccurate. 
In this section, we examine the integrated elastic energy and compare this to the
results that Persson theory posits. 

In order to test the correction factor that is present in Persson theory,
we calculate it numerically using
\begin{eqnarray}
\widetilde{S}(p_{0},\zeta) = 
  \frac{{\cal E}_{\rm exa}(p_{0},\zeta)-{\cal E}_{\rm exa}(p_{0},\zeta-\Delta \zeta)}
       {{\cal E}_{\rm P}(p_{0},\zeta)-{\cal E}_{\rm P}(p_{0},\zeta-\Delta \zeta)},
\label{eq:S_q_numerical}
\end{eqnarray}
where the numerator comprises the subtraction of the \textit{total} elastic energy 
of a simulation in which everything up to a magnification of $\zeta$ is resolved 
from that of a simulation with a slightly lower magnification. 
The denominator is the difference between \eref{eq:elasticEnergy_h_tot} 
for two different values of $\zeta = q/q_{\rm t}$, which leaves only terms
due to the \textit{newly resolved wavelengths}.
We increment the resolution
by the minimum amount possible for a given discretization, i.e. recompute the contact for each new wave number separately.
This expression would yield the correction factor if each mode of the elastic energy 
were excited exactly at its appropriate resolution and remained constant with any further 
increase of resolution. Similarly in that case, 
\eref{eq:elastic_energy_norm} would yield \eref{eq:elastic_energy_norm_predicted}.

\begin{figure}[htbp]
\includegraphics[width=\textwidth]{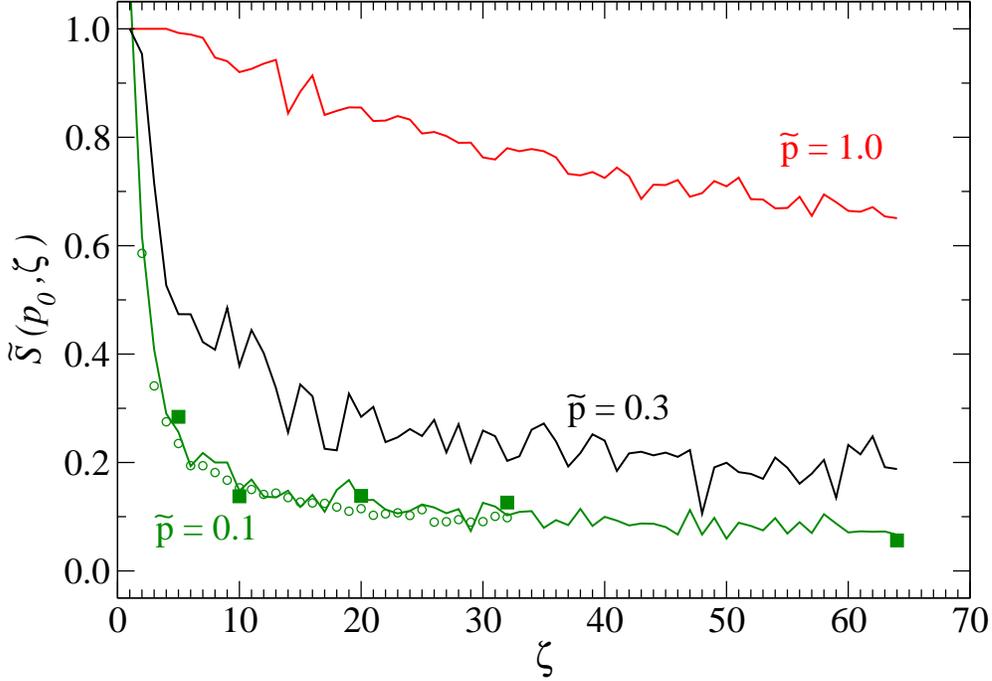}
\caption{ 
Incremental correction factor $\widetilde{S}(p_{0},\zeta)$, see \eref{eq:S_q_numerical},
versus magnification, for different pressures. 
Solid lines are $N=1024$, while open circles ($N=512$) and filled squares ($N=2048$) are 
size-scaled data which shows that the values are converged. The data is averaged over
between $10$ and $2,500$ different realizations of the rough surface.
}
\label{fig:dS}
\end{figure}

The numerator varies 
quite substantially for different values of $q$. As a consequence $\widetilde{S}(p_{0},\zeta)$ 
converges very slowly, so \fref{fig:dS} is the result of more than $2,400$ sets of 
independent random instances, for a total of about $2.9$ million simulations of size 
$512\times 512$. We confirmed the results with higher-resolution 
simulations at $N = 1024$ ($95$ sets, $\sim 400,000$ simulations), and, at selected magnifications, 
with $N=2048$ ($\sim 100$ sets, $80,000$ simulations). The latter also include magnifications up to $\zeta = 128$.

\begin{figure}[htbp]
\includegraphics[width=\textwidth]{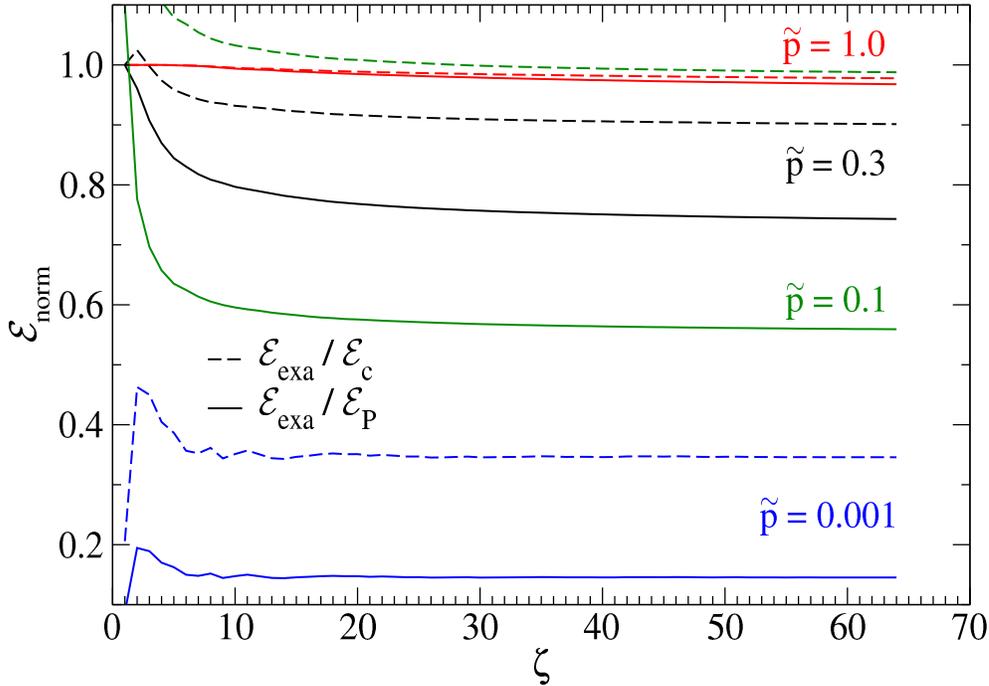}
\caption{ 
Integrated normalized elastic energy. The solid lines represent the 
original variant of Persson theory (with $S(\zeta)=1)$, while the 
dashed lines hold for $S$ as given in \eref{eq:gamma}. For high
and intermediate pressures, the corrected expression yields very good
values. At $\tilde{p}=0.1$ and low magnification, the correction factor
boosts the elastic energy beyond the measurement, while at even 
lower pressures, the result is far below the measurement.
Each data set is an average over $10$ different random realizations.
}
\label{fig:E_vs_zeta}
\end{figure}

\begin{figure}[htbp]
\includegraphics[width=\textwidth]{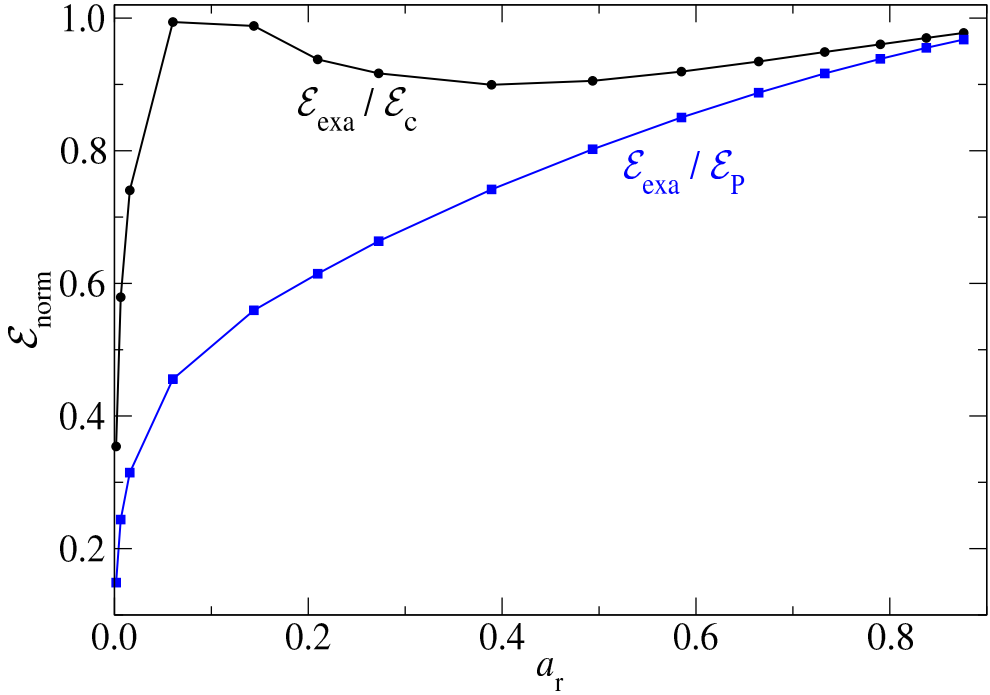}
\caption{ 
Similar to \fref{fig:E_vs_zeta}, except that the integrated normalized 
elastic energy is plotted against the relative contact area (at constant 
magnification $\zeta=64$). Each data set is an average over $10$ different random 
realizations. The correction factor works well for $a_{\rm r}\gtrsim0.05$;
at even lower pressures, the deviation from the measurement increases 
dramatically.
}
\label{fig:E_vs_ar}
\end{figure}

Since we know from \fref{fig:elastic_energy_H08} and~\ref{fig:elastic_energy_H03}
that the individual modes of the elastic energy ${\cal E}_{\rm norm}(\zeta)$ do \textit{not}
behave exactly as assumed in Persson theory, \eref{eq:S_q_numerical} may not be 
an appropriate comparison. Instead, we consider the deviations of the \textit{total} 
elastic energy with respect to the magnification and to the measured relative contact area. 
\begin{eqnarray}
{\cal E}_{\rm norm}(p_{0},\zeta) = \frac{{\cal E}_{\rm exa}(p_{0},\zeta)}{{\cal E}_{\rm P}(p_{0},\zeta)},
\end{eqnarray}
The results are shown in \fref{fig:E_vs_zeta} (varying $\zeta$ for $p_{0}={\rm const}$)
and \fref{fig:E_vs_ar} (for $\zeta=64$, varying $p_{0}$, and therefore $a_{\rm r}$).
They reveal
that using \eref{eq:gamma} indeed significantly improves agreement between theory and 
numerical measurements compared with the original theory where $S\equiv 1$. 
Nevertheless, the total elastic energy is still overestimated by $\approx 10\%$ 
even with the more complicated functional form. At low pressures, even the 
correction factor is insufficient to get theory and measurement to agree.

\section{Conclusions}\label{sec:conclusions}

In this work, we have revealed quite significant shortcomings of the
assumptions entering Persson's contact mechanics theory.
However, many errors cancel, which explains why quite a few interfacial
properties are predicted very accurately by the theory.
It is nevertheless not clear if {\it all} interfacial properties benefit
from such cancelations so that the theory might need to be improved
for some applications. 
For example, in the context of rubber friction or other problems involving
moving interfaces, the inaccurate partitioning of energy amongst different 
modes might be problematic.
At the current stage of development, only the {\it net} elastic energy of a
relaxed configuration turns out reasonable. 

Despite our criticism, we recognize Persson theory as the only theory for rough,
linearly-elastic contacts that is based on controlled approximations and reveals
accurate information not only on scalar numbers but also on distribution
functions including contact geometry.
The theory is essentially only based on directly measurable quantities and thus 
free of \textit{ad-hoc} parameters except for one correction factor of order unity.
In this work we found evidence that this correction factor is needed but has
so far been implemented only heuristically.
Our results indicate that the correction factor does not yield very accurate elastic
energies at low pressure and moreover is sensitive to the rate at which 
the relative contact area decreases with increasing resolution rather than to the 
relative contact area itself.
One reason why the gap distribution functions are nevertheless predicted quite
accurately may be that for low pressures the mean gap is approximately only logarithmic in 
pressure~\cite{Persson07PRL,Yang08JPCM,Almqvist11,ProdanovDappMueser2014a}.
As a consequence, a change in gap is only a logarithmic function of energy so
that only the order of magnitude of the energy needs to be known. For high pressures,
on the other hand, Persson theory is quite accurate.

Correcting individual ingredients to the contact mechanics theory
might disturb a delicate balance of error cancelations which is currently present.
It might therefore be necessary to make adjustments to the elastic energy
and the various aspects of the diffusion analogy simultaneously so that the
predictions of the quantities tested so far do not deteriorate. 

It would certainly be desirable to motivate corrections to the theory 
without making {\it ad-hoc} or uncontrolled assumptions.
One possible avenue to derive such corrections is to consider a model
in which hard-wall repulsion is replaced with smoother repulsion, such
as exponential repulsion, which is more amenable to analytical calculations
than (non-holonomic) hard-wall constraints. 
In fact, one of the authors investigated such a model~\cite{Muser08PRL}
and proposed that stress-dependent drift arises
beyond a second-order cumulant expansion of the model,
which is formally equivalent to Persson theory. 
Pursuing such an expansion might, however, turn out tedious.
The analytical expressions become increasingly involved with each added order 
and the next non-vanishing term for colored-noise surfaces only arises 
at fourth order. 
%

\ack
We thank the J\"ulich Supercomputing Centre for computing time on JUQUEEN and
JUROPA. MHM also thanks DFG for support through grant No. Mu 1694/5.

\appendix

\section{On-the-fly determination of the diffusion coefficient}
\label{sec:onTheFly}

The deduction of drift and diffusion coefficients from data
for the transition probability, 
${\rm Pr}(p,\zeta+\Delta\zeta\vert p',\zeta)$, 
as shown in \fref{fig:broadening},
becomes non-trivial when the magnification-induced broadening
is no longer small compared to $p'$.
It is then no longer sufficient to assume that the increase of the
second moment of the pressure from old to new magnification reflects
the broadening.
An example is the case for the distribution
associated with the bin number 10 in \fref{fig:broadening}.
The reason why looking at the change in the second moment of the distribution
is no longer sufficient is that some walkers belonging to bin 10 at the original
magnification have fallen out of contact at the new magnification, as one
can see, in \fref{fig:broadening}, by the (blue) triangle at $p=0$.
This is why walkers having landed at $p=0$ no longer contribute to the 
random walk, at least as long as they stay outside the contact. 
Thus, rather than fitting to Gaussians, it would be better to fit
the new probabilities
to the function displayed on the r.h.s. of \eref{eq:transitionFinal}.
The value for $p'$, however, would be allowed to differ from the first moment
of the pressure associated with the original bin. 

Instead of fitting the measured transition probabilities to 
\eref{eq:transitionFinal}, we ask the question what parameters
$p'$ and $\Delta p$ should be used to reproduce the first two
moments of the individual bin distribution at the new magnification.
It can be readily shown that the first moment of the distribution
is identical to $p'$.
As a consequence, we can simply set  
\begin{equation}
p' = \langle p \rangle_{n,\zeta+\Delta \zeta},
\end{equation}
where $\langle \bullet \rangle_{n,\zeta+\Delta \zeta}$ indicates an average
over all walkers in bin $n$ at the new magnification.
As a consequence, the drift in pressure can be computed from the 
difference of the first moments at two consecutive magnifications, i.e., 
\begin{equation}
\mu_n = \frac{\langle p \rangle_{n,\zeta+\Delta\zeta} -
\langle p \rangle_{n,\zeta} } {\Delta\zeta}.
\end{equation}

In a similar fashion as done for the first moment, we can equate the 
second moment of the pressure as obtained in the simulation and 
as deduced by the distribution function via
\begin{eqnarray}
\fl\langle p^2 \rangle_{n,\zeta+\Delta\zeta} &= 
\frac{1}{\sqrt{2\pi\Delta p_n^2}}
\int_0^\infty {\rm d}p \; p^2  \;
\left[ \exp\left\{- \frac{(p-p')^2}{2\Delta p^2_n}\right\}  \right.
\nonumber\\ &
\;\;\;\;\;\left.
-  \exp\left\{- \frac{(p+p')^2}{2\Delta p^2_n}\right\} \right]
\nonumber\\
&=
\sqrt{\frac{2}{\pi}}\; p'\;\Delta p_n \exp\left\{-\frac{p'^2}{2\Delta p_n^2}\right\} \
\nonumber\\ 
&\;\;\;\;\;+ 
\left(p'^2+\Delta p_n^2\right) {\rm erf }\left(\frac{p'}{\sqrt{2}\Delta p_n} \right).
\label{eq:2ndMoment} 
\end{eqnarray}
While \eref{eq:2ndMoment} cannot be inverted analytically to 
solve for $\Delta p_n$, we found that 
\begin{equation}
\frac{\Delta p_n}{p'} \approx \sigma_n
\frac{1+\alpha_1 \sigma_n + \alpha_2 \sqrt{\frac{\pi}{8}} \sigma_n^2}
{1+\alpha_2\sigma_n}
\label{eq:approxDeltaPress}
\end{equation}
with $\alpha_1 = 0.5153$, $\alpha_2 = 0.7591$, and
\begin{equation}
\sigma^2_n = 
 \frac{\langle p^2 \rangle_{n,\zeta+\Delta}}{p'^2}  - 1 
\end{equation}
is exact in the 
limits of $\sigma_n \to 0$ and $\sigma_n \to \infty$
and yields results with errors less than 3\% in between those limits. 
If higher accuracy is needed, one may use \eref{eq:approxDeltaPress}
as a starting point for a Newton's method. 

In this calculation, we have neglected that the pressure distribution in the initial
bin is not an exact delta function but has a finite width $\delta p_n$ that is 
typically in the order of but smaller than the half width of the bin itself.
This induces a (small) artifical broadening of the final distribution function, 
which can be accounted for by replacing 
$\Delta p_n^2$ with $\Delta p_n^2-\delta p_n^2$. 
With this new $\Delta p_n$, one can compute the diffusion coefficient 
associated with bin $n$ according to
\begin{equation}
D_n = \frac{\Delta p_n^2}{\Delta \zeta}.
\end{equation}

\section*{References}
\providecommand{\newblock}{}

\end{document}